\documentclass[12pt]{iopart}
\usepackage{amssymb}
\usepackage{epsf}  

\begin{document}

\title{Medium Modification of the Jet
Properties}

\author{Carlos A. Salgado}

\address{Department of Physics, CERN, Theory Division CH-1211 Geneva 23}

\begin{abstract}
In the case that a dense medium is created in a heavy ions collision,
high-$E_t$ jets are expected to be broadened by
medium-modified gluon emission. This broadening is directly related, 
through geometry, to the
energy loss measured in inclusive high-$p_t$ particle suppression.
We present here the modifications of jet observables due to the
presence of a medium for the case of azimuthal jet energy distributions and
$k_t$-differential multiplicities inside the jets.
\end{abstract}



%\maketitle

Recent data from RHIC on inclusive particle production at high-$p_t$
are well described in terms of radiative parton energy loss~\cite{vitev}. 
Associated to this energy loss, theory predicts
a broadening of the radiated gluon spectrum when compared with the 
vacuum case \cite{baier,nosjet}. Indeed, the average transverse momentum
of the radiated gluons is given by $\langle k_t^2\rangle \sim
\omega_c/L$, where $\omega_c$ is the critical energy above which the 
medium-induced gluon radiation is suppressed by formation time effects (see
below). 

The general expression for the radiation of a gluon with energy $\omega$ and
transverse momentum $k_t$ in the limit of large quark energy takes the form
~\cite{Wiedemann:2000za}
\begin{eqnarray}
  \omega\frac{dI}{d\omega\, d{\bf k}}
  &=& {\alpha_s\,  C_R\over (2\pi)^2\, \omega^2}\,
    2{\rm Re} \int_{0}^{\infty}\hspace{-0.3cm} dy_l
  \int_{y_l}^{\infty} \hspace{-0.3cm} d\bar{y}_l\,
   \int d^2{\bf u}\,
  e^{-i{\bf k}_t\cdot{\bf u}}   \,
  e^{ -\frac{1}{2} \int_{\bar{y}_l}^{\infty} d\xi\, n(\xi)\, \sigma({\bf u})}\,
  \times
\nonumber\\
  &\times&{\partial \over \partial {\bf y}}\cdot
  {\partial \over \partial {\bf u}}\,
  \int_{{\bf y}={\bf r}(y_l)}^{{\bf u}={\bf r}(\bar{y}_l)}
  \hspace{-0.5cm} {\cal D}{\bf r}
   \exp\left[ i \int_{y_l}^{\bar{y}_l} \hspace{-0.2cm} d\xi
        \frac{\omega}{2} \left(\dot{\bf r}^2
          - \frac{n(\xi)\, \sigma({\bf r})}{i\,2\, \omega} \right)
                      \right]\, .
    \label{eqspec}
\end{eqnarray}

We have studied this formula in two approximations. First, 
in the saddle point approximation, $\sigma({\bf r})=C{\bf r}^2$, 
the path integral of (\ref{eqspec}) is that of a 
harmonic oscillator of imaginary frequency. This approximation
corresponds to multiple soft scatterings, and neglects the high-$p_t$ tails
of single scattering centers. Second, in the opacity expansion, the integrand
of (\ref{eqspec}) is expanded in powers of 
$(n(\xi)\, \sigma({\bf r}))^N$. This corresponds to a fixed
number of $N$ scattering centers. In the following, the label
{\it single hard scattering} denotes the first
term ($N=1$) in this expansion.

The medium-induced gluon radiation spectrum can be written in terms of
two scaling variables \cite{nosjet},

\begin{equation}
\kappa^2=\frac{k_t^2}{\hat qL} \ , \ \omega_c=\frac{1}{2} \hat qL^2
\hspace{0.5cm} \Bigg[
\kappa^2=\frac{k_t^2}{\mu^2} \ , \ \omega_c=\frac{1}{2} \mu^2L
\ \ {\rm for\ single\ hard}\Bigg]\ ,
%\hat q\rightarrow \frac{\mu^2}{L}
\label{eqvar}
\end{equation}

\begin{figure}
\begin{center}
\epsfxsize=9cm
\epsfbox{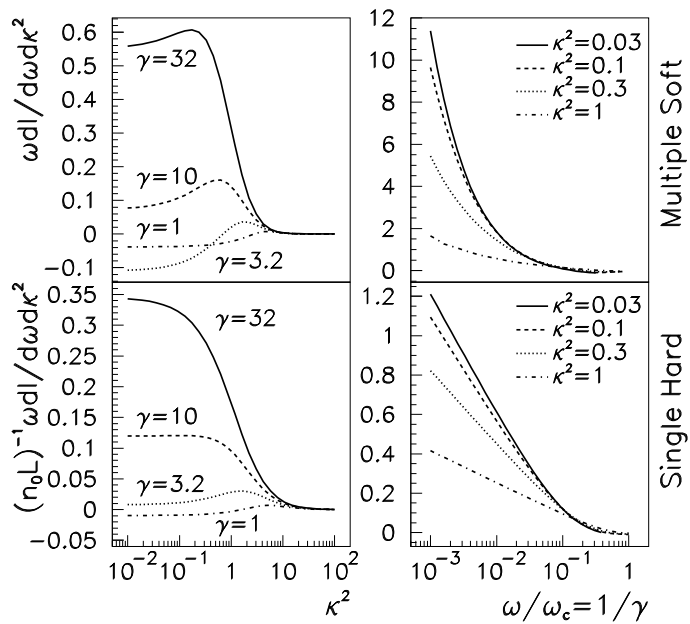}
\end{center}
\caption{\label{salgfig1}
The energy distribution of gluons radiated by a quark
as a function of the rescaled gluon energy
$\omega/\omega_c$ and the rescaled gluon transverse momentum
$\kappa$, see eq. (\ref{eqvar}), with $\alpha_s$=1/3.}
\end{figure}

\noindent
In the multiple soft scattering approximation, the only parameter is
the transport coefficient, $\hat q$. This quantity parametrizes the 
average momentum transfer squared and is proportional to the density 
of the medium. In the single hard scattering, the Debye screening mass 
$\mu$ denotes the average momentum transfer per scattering center.
In Fig. \ref{salgfig1} 
we present the double differential spectrum (\ref{eqspec}) in $\omega$ and 
$k_t^2$.

Qualitative properties of the
medium-induced gluon radiation spectrum can be understood by
coherence arguments \cite{meu,nosohq}. The relevant phase for gluon emission is
\begin{equation}
  \varphi = \Bigg\langle \frac{k_t^2}{2\omega}\, \Delta z \Bigg\rangle
%\Longrightarrow
%l_{coh}\sim\frac{2\omega}{k_t^2}
\hskip 0.5cm 
\Longrightarrow\hskip 0.5cm 
\varphi\sim\frac{L}{l_{coh}}
\sim \frac{\kappa^2}{\omega/\omega_c}\ \ {\rm for}\ \   \Delta z\sim L.
\end{equation}
\noindent
Gluon radiation occurs if the accumulated phase $\varphi\gtrsim 1$.
Thus, the radiation is suppressed for $\kappa^2\lesssim\omega_c/\omega$ and
for $\omega\gtrsim \omega_c$ (as $\kappa\lesssim 1$). 
These features are observed in the numerical
calculations presented in Fig. \ref{salgfig1}. The presence of a
limiting energy $\omega_c$ implies that the average energy loss
$\Delta E\sim\omega_c\propto L^2$, as first pointed out in 
\cite{bdmps}.

The $k_t$-integrated spectrum (see Refs. \cite{nos2,nosjet})
is also suppressed for small
values of $\omega$. This fact can also be understood by formation
time arguments:  the integration limit $k_t < \omega$ cuts the 
gluon energies $\omega^2\lesssim k_t^2$, on the other hand, the
spectrum is suppressed for $k_t\lesssim \hat q L$ due to formation
time effects. This suppresses the spectra for $\omega/\omega_c\lesssim 
\sqrt{2/\omega_cL}$. 
In this way,
formation time effects make the spectrum stable in the infrared region
\cite{nos2}. 
% This
% has important consequences for the experimental observables as we will see.

Based on this formalism of medium-induced parton energy loss, 
we have computed the medium-modification of two jet observables,
namely the fraction $\rho(R)$ of jet energy inside a cone $R$ and
the gluon multiplicity distribution~\cite{nosjet}.

\begin{figure}[h]\epsfxsize=8.5cm
\centerline{\epsfbox{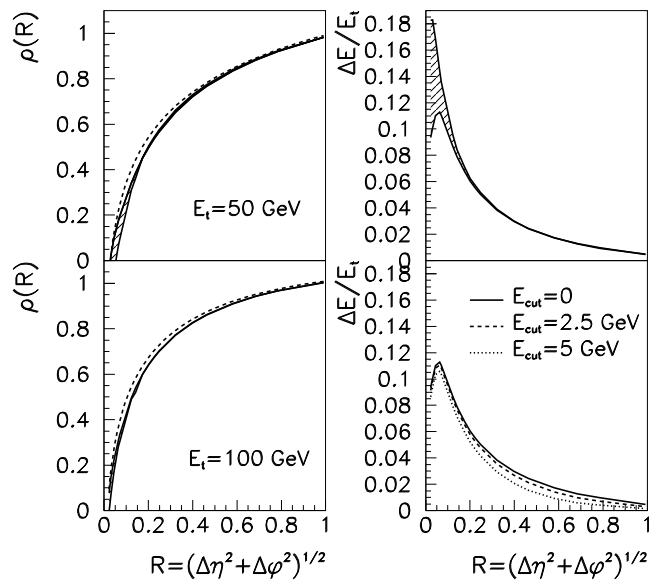}}
%\vspace{0.5cm}
\caption{LHS: The jet shape (\protect\ref{eq5}) for a 50 GeV and
100 GeV quark-lead jet which fragments in the vacuum (dashed curve) or
in a dense QCD medium (solid curve)
characterized by $\omega_c = 62$ GeV and $\omega_c\, L = 2000$.
RHS: the corresponding
average medium-induced energy loss for $E_t = 100$ GeV
outside a jet cone $R$ radiated
away by gluons of energy larger than $E_{\rm cut}$. Shaded regions
indicate theoretical uncertainties discussed in the text.
}\label{figfig2}
\end{figure}

The fraction of the jet energy inside a cone of radius
$R=\sqrt{(\Delta \eta)^2 + (\Delta \Phi)^2}$ is
\begin{equation}
  \rho_{\rm vac}(R) = \frac{1}{N_{\rm jets}} \sum_{\rm jets}
  \frac{E_t(R)}{E_t(R=1)}\, .
  \label{eq5}
\end{equation}
\noindent
In the presence of the medium, this energy is shifted by~\cite{nosjet}
\begin{equation}
  \rho_{\rm med}(R) =
    \rho_{\rm vac}(R) - \frac{\Delta E_t(R)}{E_t}
        + \frac{\Delta E}{E_t} \left( 1 - \rho_{\rm vac}(R)\right)\, .
      \label{eq9}
\end{equation}
Here, $\Delta E_t(R)$ is the additional (medium) energy radiated
outside a cone $\Theta=R$,
$\Delta E(\Theta)=\int \epsilon P(\epsilon,\Theta)d\epsilon$. It
is determined by the probability $P(\epsilon,\Theta)$ that an energy
fraction $\epsilon$ is radiated outside $\Theta$, which can be calculated
from eq.~(\ref{eqspec}) \cite{nos2}.
In Fig.~\ref{figfig2} we plot the medium-shifted distribution (\ref{eq9}). 
The shaded area corresponds to the uncertainty in finite quark-energy 
effects: in the eikonal approximation
$P(\epsilon)$ has support in the unphysical region $\epsilon > 1$.
To estimate this effect  we take
$P(\epsilon)\,\rightarrow\, P(\epsilon)/\int_0^1 d\epsilon P(\epsilon)$.

%\noindent
The effect of the medium is very small (at $R$=0.3, $\sim$ 5\% for
a 50 GeV jet and $\sim$ 3\% for a 100 GeV jet). The smalleness of this effect
could allow for a calibration of the total energy of the jet
without tagging a recoiling hard photon or Z-boson. It also implies
that the jet $E_t$ cross section scale with the number of binary collisions.
In order to check the sensitivity of our results to the small-energy region,
we impose low momentum cut-offs which remove gluon emission below 5 GeV. 
As anticipated, the
transverse broadening is very weakly affected by these cuts. This is due to
the infrared behavior of the spectrum for small values of $\omega$.
A proper subtraction of the large background present
in heavy ion collisions would benefit from this result.

\begin{figure}[t!]\epsfxsize=8cm
\centerline{\epsfbox{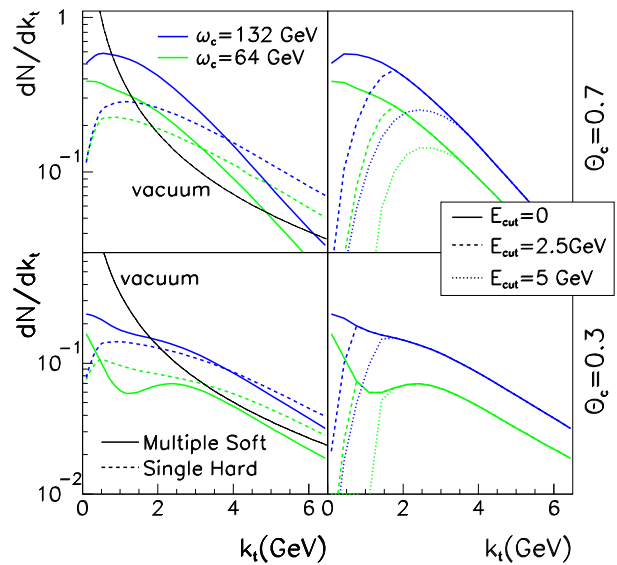}}
\caption{Comparison of the vacuum and medium-induced part of the
gluon multiplicity distribution
(\protect\ref{eq8}) inside a cone size $R=\Theta_c$, measured as a
function of $k_t$ with respect to the jet axis. 
%Removing gluons with
%energy smaller than $E_{\rm cut}$ from the distribution (dashed and dotted
%lines) does not affect the high-$k_t$ tails.
}\label{figglmult}
\end{figure}

$k_t$-differential measurements are expected to be more sensitive to
medium effects. We studied the intrajet multiplicity of produced gluons as
a function of the transverse (with respect to the jet axis) momentum.
The medium-induced additional number of gluons
with transverse momentum $k_t = \vert {\bf k}\vert$, produced
within a subcone of opening angle $\theta_c$ is
\begin{eqnarray}
 \frac{dN_{\rm med}}{dk_t} =  \int_{k_t/\sin\theta_c}^{E_t} d\omega\,
               \frac{dI_{\rm med}}{d\omega\, dk_t}\, .
  \label{eq8}
\end{eqnarray}
For the vacuum we simply assume
$dN_{\rm vac}/dk_t\sim 1/k_t\log(E_t\sin\theta_c/k_t)$.
The total multiplicity is the sum of the two.
Fig. \ref{figglmult} shows that this multiplicity distribution is 
very sensitive to medium-effects since it broadens significantly. 
Moreover, the high-$k_t$ tail of the distribution is unaffected by
the soft background since only high-$p_t$ particles can have a large
$k_t$ {\it inside} the jet cone.

%The effect is now rather large for transverse momemta of
%the order of several GeV. So, we expect that this observable could give
%very valuable experimental information about the dynamics of the
%medium--induced gluon radiation and, hence, about the medium.
%A more realistic analysis, however,
%would need to include the whole fragmentation, etc...
%Nevertheless, the origin of the shift to larger transverse momentum is mainly
%kinematic, understandable by qualitative arguments. In this
%way, we expect this conclusion to be very robust and not depending on the
%actual realization of the model.

The first results on 'jet-like' particle production associated to high-$p_t$
trigger particles have been presented at this conference \cite{jetsqm}. The 
results are in rough qualitative agreement with the ones shown 
here for much larger jet energies. A more refined analysis of finite energy
corrections is, however, needed for a direct comparison with data.

\end{document}